\documentclass[10pt,letterpaper]{article}
\usepackage{opex3}
\usepackage{amsmath,amssymb}
\usepackage{mathrsfs}
\usepackage{graphicx}
\usepackage{dcolumn}
\usepackage{bm}

\begin{document}

\title{Remnants of semiclassical bistability in the few-photon regime of cavity QED}
\author{Joseph Kerckhoff,$^{1,2\ast}$ Michael A. Armen,$^1$ and Hideo Mabuchi$^1$} 
\address{$^1$Edward L.\ Ginzton Laboratory, Stanford University, Stanford, California 94305, USA\\
$^2$Current address: JILA, University of Colorado and NIST, Boulder, Colorado 80309, USA}
\email{$^\ast$jkerc@jila.colorado.edu}

\date{\today}

\begin{abstract}
Broadband homodyne detection of the light transmitted by a Fabry-Perot cavity containing a strongly-coupled $^{133}$Cs atom is used to probe the dynamic optical response in a regime where semiclassical theory predicts bistability but strong quantum corrections should apply. While quantum fluctuations destabilize true equilibrium bistability, our observations confirm the existence of metastable states with finite lifetimes and a hysteretic response is apparent when the optical drive is modulated on comparable timescales. Our experiment elucidates remnant semiclassical behavior in the attojoule ($\sim10$ photon) regime of single-atom cavity QED, of potential significance for ultra-low power photonic signal processing.
\end{abstract}


\noindent Research over the past decade in single-atom cavity quantum electrodynamics (cQED) has largely focused on the generation of non-classical states of light and on the development of devices for quantum information protocols~\cite{Obri09,Aoki09,DiCa10}. However single-atom cQED with strong coupling is also a natural context in which to study the interplay of nonlinear mean-field dynamics and quantum fluctuations~\cite{Sava88,Remp91,Arme06,Kerc11a}, which is a topic of equally broad fundamental interest and with potentially greater near-term relevance for information technology \cite{Mill10}. Dynamical systems theory~\cite{Stro01} provides key methods and insights for nonequilibrium statistical mechanics, as well as essential tools for analyzing and engineering nonlinear phenomena. Substantial work will be required to generalize these methods to encompass quantum coherence and fluctuations for applications in the emerging disciplines of nanoscale and quantum engineering.  Absent a systematic understanding of quantum stochastic nonlinear phenomena in the strong coupling regime, intuitive connections to semiclassical theory provide important guidance for analyzing and engineering the behavior of open quantum systems \cite{Arme06,Mabu11}.  Here we show that the optical response of a strongly driven single-atom cQED system clearly displays competing influences of semiclassical bistability and quantum fluctuations, in quantitative agreement with theory.

Our experiment utilizes a single gas-phase $^{133}$Cs atom as the nonlinear medium in a Fabry-Perot optical resonator, but we note that the same physics should also be relevant to quantum nonlinear dynamics in strongly coupled nanophotonic systems~\cite{Srin07, Fara08}. We study the dynamic input-output properties of the atom-cavity system in a parameter regime that semiclassically~\cite{Lugi84} would be expected to exhibit absorptive bistability with a hysteresis loop~\cite{Szok69} suitable for attojoule ($\sim10$ photon) optical switching~\cite{Smit86}. Theoretical studies have shown~\cite{Sava88,Kili91,Arme06} however that quantum fluctuations of the cavity field and atomic dipole should induce spontaneous switching between low- and high-transmission states, destroying true optical bistability, when the energy separation of the semiclassical dynamical attractors reaches the few-photon scale. Previous experimental studies have proven the need to utilize fully quantum models for predicting the steady-state optical response in single-atom cQED with strong driving~\cite{Hood98}. Here we take an important step further by recording dynamic signals that reaffirm the quantum mechanical model, but nonetheless reveal remnant behavior reminiscent of the semiclassical picture even in the deep quantum regime. As research on nanophotonic logic devices pushes towards attojoule-scale switching energies~\cite{Yang07,Noza10,Mill10}, functionalizing such remnant behaviors will become a definitive challenge.

\begin{figure*}[t]
\begin{center}
\includegraphics[width=1\textwidth]{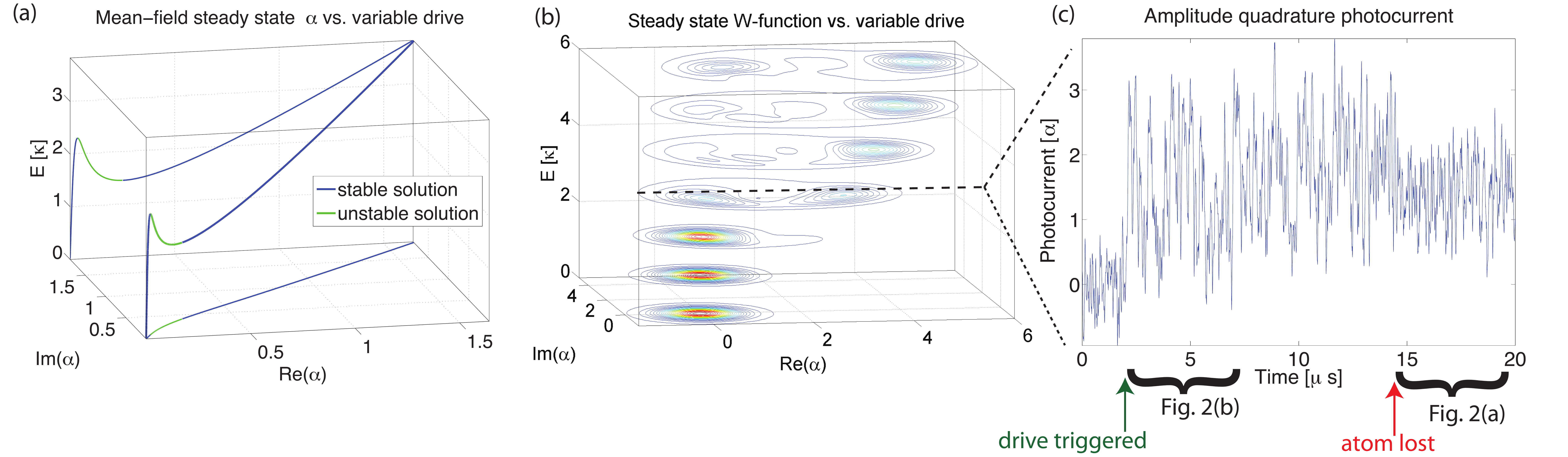}
\end{center}
\caption{\label{fig:MF} (a) Steady state intracavity mode amplitude $\alpha$ from the semiclassical Maxwell-Bloch Equations (MBEs) as a function of drive amplitude $E$ for the experimental cQED system with $\{\Theta,\Delta\} = \{-1.1,.7\}\kappa$ mode-drive and atom-drive detunings. Blue (green) indicates dynamically stable (unstable) solutions, predicting true amplitude bistability for drives in the interval $E = [1.5,2.3]\kappa$. (b) Wigner quasi-probability functions of the steady state cavity field from the quantum model as a function of drive amplitude for the same parameters as (a). (c) Sample trace of amplitude quadrature homodyne measurement of the transmitted field during an atom transit with the drive turned on at $t=2\mu$s and held at $E=2.6\kappa$. The slight decrease in signal variance between 2 and 14$\mu$s is likely due to a gradual decrease in the coupling rate $g$ from a initial, near-maximal value as the atom moves through the cavity mode.}
\vspace{-0.1in}
\end{figure*}

The Maxwell-Bloch Equations (MBEs) represent a semiclassical mean-field approximation to the cQED master equation (see Appendix \ref{sec:models}) in the limit of weak coupling and large atom number~\cite{Lugi84,Carm86} and have been used in research on optical bistability with atomic ensembles as the intracavity nonlinear medium~\cite{Gang90a,Gang90b,Arme06}. In single-atom cavity QED with strong coupling and saturating driving fields, the MBEs should not apply {\it a priori} as they ignore correlations between the atom and intra-cavity photons. However, the MBEs do retain some relevance as a projection of the quantum master equation onto the sub-manifold of semiclassical atom-field states~\cite{Mabu08a} and numerical studies have shown~\cite{Arme06} that solutions of the full quantum model often mimic those of the MBEs qualitatively.

The MBEs predict absorptive optical bistability for the parameters of our experiment, as shown in Fig.~\ref{fig:MF}(a). The vertical axis represents input drive amplitude (with $\sqrt{27\text{pW}}$ per unit on the vertical axis), and the solution curve passes through horizontal planar coordinates propotional to the real and imaginary parts of the output steady-state complex field amplitude(s).  The projection of the curve onto the bottom face of the coordinate box indicates the phases of the steady-state solutions (relative to the optical phase of the laser drive), while the projection onto the back face illustrates the input-output relation for amplitude-quadrature homodyne detection.  In this representation, a characteristic `S-curve' appears with two stable solutions and one unstable solution co-existing for $E$ in the range of $[1.5,2.3]\kappa$. Using identical parameters, the steady-state solution of the quantum master equation can be determined for each value of the drive strength. Contours of the corresponding intracavity field Wigner functions (tracing over the atomic states) are displayed for several drive strengths in Fig.~\ref{fig:MF}(b).  A double-peaked structure emerges along the amplitude quadrature in a range of $E/\kappa$ similar to the bistable region of the MBEs.  The bimodal steady states correspond to an incoherent mixture of two states: a weakly excited atom and low-amplitude field state, and a fully saturated atom and a high-amplitude field state.  Thus, the two peaks of the distribution may be qualitatively associated with the low- and high-amplitude branches of the semiclassical absorptive bistability curve, but in the quantum model neither is truly stable \cite{Sava88}.  It can be seen that in any single trial the cavity field (and thus the output power) spontaneously switches between low and high states, as in the experimental data of Fig.~\ref{fig:MF}(c) (described below). This stochastic switching is a dramatic consequence of quantum fluctuations in this non-linear, attojoule-scale optical system.

In our experiment \cite{Kerc11a,Arme09}, laser cooled $^{133}$Cs atoms are dropped into a Fabry-Perot optical resonator (length 27$\mu$m, 10cm radius of curvature mirrors, field decay rate $\kappa/2\pi =$ 9.3MHz) supporting a circularly-polarized, 852nm TEM$_{00}$ mode actively frequency-stabilized relative to the $(6S_{1/2},F=4,m_F=+4)\rightarrow(6P_{3/2},F=5,m_F=+5)$ atomic cycling transition (with dipole decay rate $\gamma_\perp/2\pi = 2.6$MHz). As an atom falls through the cavity mode it experiences a position-dependent coupling rate $g$ (maximum $g_0/2\pi =$56.8MHz at the cavity anti-nodes) that can be monitored via the transmission of a weak and detuned circularly-polarized optical probe. Once a strongly coupled atom is detected, the probe power and detuning are adjusted for optical homodyne detection and 200MS/s data acquisition (first arrow at $2\mu$s on the time axis in Fig.~\ref{fig:MF}(c).  Fig.~\ref{fig:MF}(c) depicts a representative signal (all data has been post-filtered at 20MHz bandwidth for clarity); with the drive amplitude held at a fixed value of $E=2.6\kappa$, the amplitude quadrature of the transmitted field fluctuates with a large variance until $14\mu$s (red arrow) when the atom is abruptly lost and and the measured transmission settles to the shot noise-variance signal with intermediate mean amplitude expected for our cavity when empty (perturbed only by slight $\sim$1-10kHz mechanical instabilities).  See Appendix \ref{sec:exp} for more explanation of the apparatus.

\begin{figure*}[t]
\begin{center}
\includegraphics[width=1\textwidth]{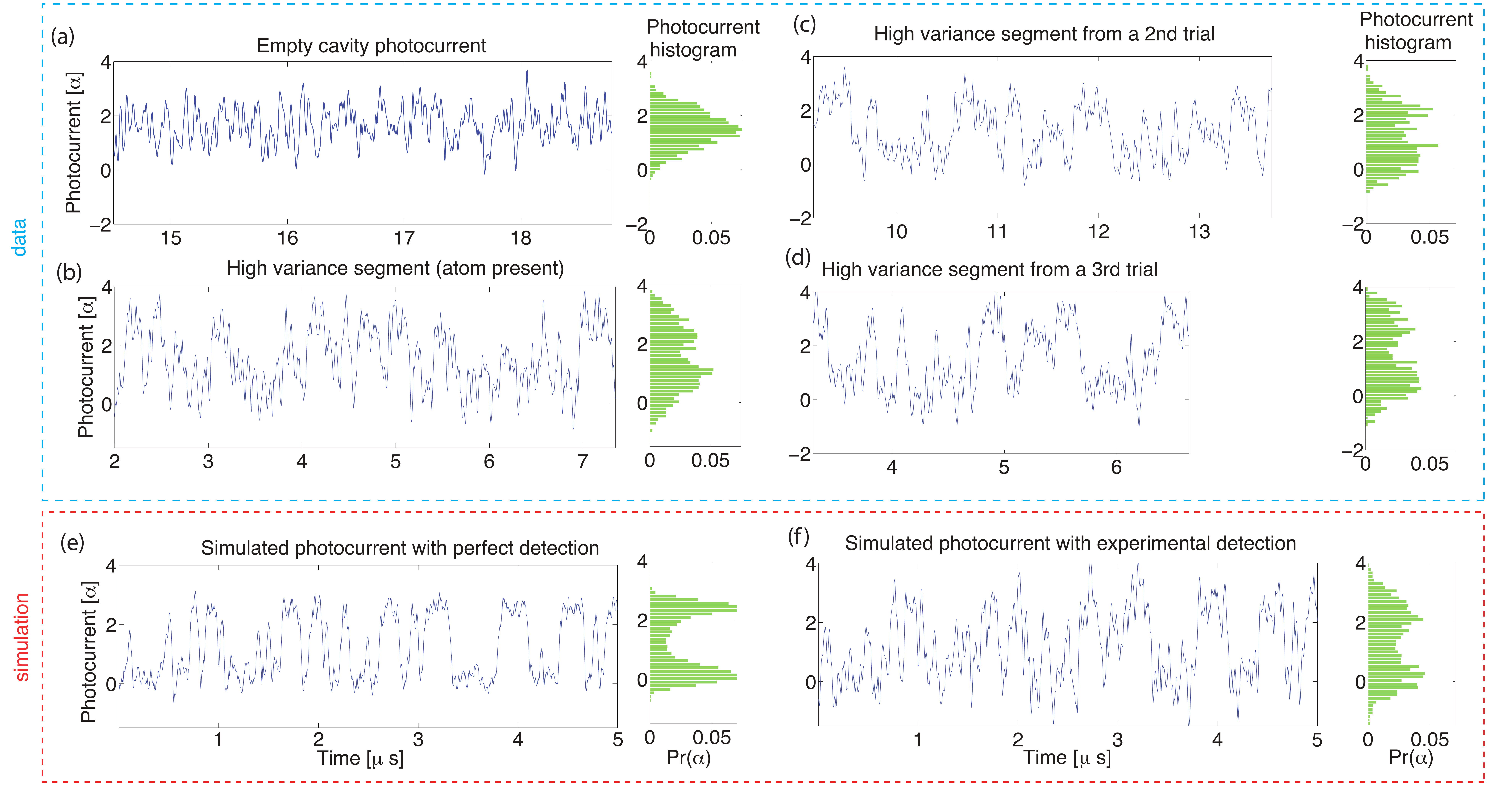}
\end{center}
\caption{\label{fig:Singles} (a) Amplitude quadrature photocurrent data segment taken from the $14-19\mu$s interval in Fig.~\ref{fig:MF}(c).  The histogram of the data points in this segment are presented on the right, revealing a distribution that well-fits the expected normal distribution of photocurrents when our cavity is empty.  (b) High-variance photocurrent segment taken from the $2-7\mu$s interval in Fig.~\ref{fig:MF}(c), when the atom should be near-maximally coupled to the cavity mode.  Faint, but sharp transitions between high and low outputs and a seemingly bimodal distribution are visible in this single shot measurement.  (c-d) Two more high variance segments from two different experimental runs with the same parameters.  (e) A typical 5$\mu$s amplitude quadrature segment simulated using quantum trajectory techniques, assuming perfect detection efficiency, with a clearly bimodal output.  (f)  The same simulated realization as (e), but with calibrated detection inefficiency yields a signal that resembles (b-d) in both visibility and time scale for the large fluctuations.}
\vspace{-0.1in}
\end{figure*}

Zooming in on segments of the trace in Fig.~\ref{fig:MF}(c) reveals significant dynamics on $\mu$s time scales.  While the signals after the atom has left the cavity have white noise, Gaussian statistics as expected (e.g. the $14-19\mu$s segment from Fig.~\ref{fig:MF}(c) is shown in Fig. \ref{fig:Singles}(a)) single shot signals observed when the drive is first turned on are more volatile (the $2-7\mu$s segment from Fig.~\ref{fig:MF}(c) is shown in Fig. \ref{fig:Singles}(b)), with suggestive, sharp transitions between low and high outputs and a seemingly bimodal distribution.  Figs.~\ref{fig:Singles}(c-d) depict two more high variance segments from two more experimental trials.  The significance of these observations may be understood from simulated measurement signals using quantum trajectory techniques (see Appendix \ref{sec:models}).  Fig.~\ref{fig:Singles}(e) shows a simulated amplitude quadrature measurement segment, assuming perfect detection efficiency of all photons that decay from the cavity.  This simulated signal randomly switches between two, roughly Gaussian-distributed output states, one with a low mean and one with a high mean.  This signal's bimodal distribution may be directly related to the bimodal, steady state Wigner function for the intracavity field calculated in Fig.~\ref{fig:MF}(b) (discussed futher below).  Re-simulating the same realization of the signal accounting for the calibrated overall photon collection efficiency (20\% in the particular data set from which traces in Fig.~\ref{fig:Singles}(a-d) were produced), yields a signal in which binary switching is much less distinct in Fig.~\ref{fig:Singles}(f), but resembles the single shot data shown in Figs.~\ref{fig:Singles}(b-d) in both the visibility and apparent time scales of the large fluctuations.

\begin{figure}[tb]
\begin{center}
\includegraphics[width=0.85\textwidth]{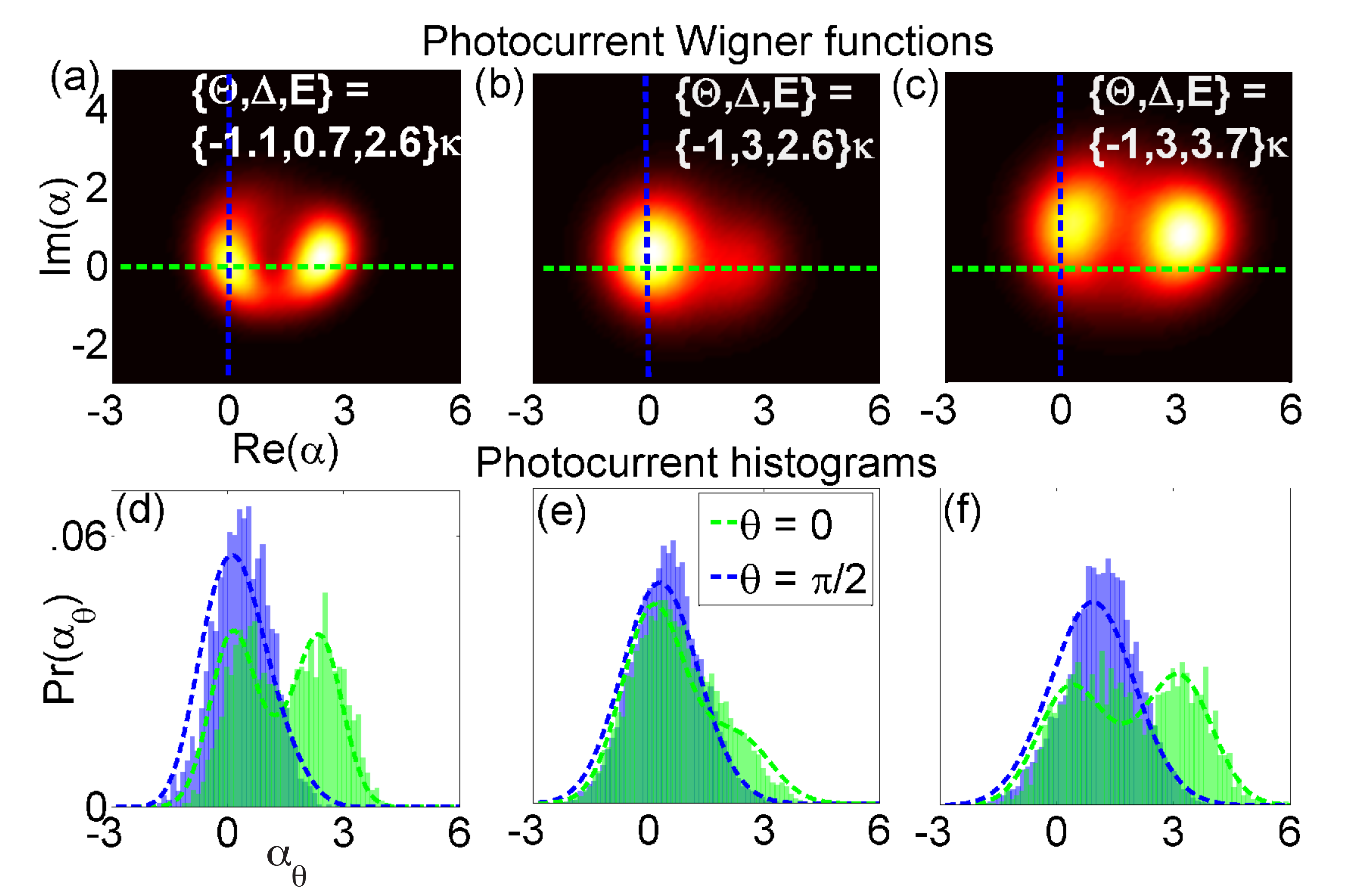}
\end{center}
\caption{\label{fig:SS}  (a-c) Wigner function representations of the expected steady state photocurrent distributions, using calibrated detection efficiencies and bandwidths, for three sets of detuning and drive parameters. Marginal distributions of homodyne measurements of any quadrature may be obtained by integrating these representations over the perpendicular quadrature (see Appendix \ref{sec:models}). (d-f) Histograms represent the phase- ($\alpha_{\pi/2}$) and amplitude-quadrature ($\alpha_0$) photocurrent distributions from ensembles of the highest-variance segments. The histograms are compared with theoretically expected distributions obtained from corresponding Wigner functions (a-c), respectively.}
\vspace{-0.1in}
\end{figure}

To confirm the steady state quantum mechanical model in the high detection bandwidth regime more quantitatively, in Fig.~\ref{fig:SS} we compare amplitude- and phase-quadrature photocurrent distributions obtained from three different sets of mode-drive, atom-drive detunings and drive amplitudes ($\Theta$, $\Delta$, and $E$, respectively), with $\sim$15\% overall cavity photon detection efficiency (independently calibrated for each parameter set, with variations attributed to thermal drifts in the signal-local oscillator optical mode matching).  Each histogram is computed from an aggregate of several few-$\mu$s-long segments of the highest variance photocurrent data, such as those presented in Figs.~\ref{fig:Singles}(b-d).  While enhanced amplitude noise is always visible when the drive is first triggered, a minority of events exhibit fluctuations of near-maximal magnitude since our atom detection scheme is triggered by strongly coupled atoms ($C\equiv g^2/2\kappa\gamma_\perp\gg1$) that may still be significantly less than near-maximally coupled; a maximum of $C=67$ is achievable with our apparatus (see Appendices \ref{sec:exp} and \ref{sec:models}).  Because of this post-selection to isolate near-maximally coupled atom transits, useful segments of photocurrent are rather sparse in our overall data set.  We are thus limited to the presentation of small number statistics, but given the very distinctive features of the bistability-related signals (to be discussed below) and the straightforward nature of our selection criterion (highest atom-induced photocurrent variance), we are confident that our analysis procedure enables us to draw physically meaningful conclusions. 

For a near-detuned system at the onset of `bistability' with $\{\Theta,\Delta,E\} = \{-1.1,.7,2.6\}\kappa$, wide/bimodal and narrow/normal distributions are apparent in the amplitude- and phase-quadrature distributions, respectively (Fig.~\ref{fig:SS}(d)). However, when the atom-drive detuning is increased to $\Delta = 3\kappa$ for the same drive amplitude, the low-amplitude transmitted field dominates (Fig.~\ref{fig:SS}(e)) as the drive threshold for `bistability' increases with $\vert\Delta\vert$. The bimodal amplitude distribution reemerges when the drive amplitude is increased to $E = 3.7\kappa$ in Fig.~\ref{fig:SS}(f). 

The data are in agreement with quantum theoretical predictions (see Appendix \ref{sec:models}), despite the use of a somewhat idealized model that assumes a static coupling rate $g$.  Whereas $g$ actually depends upon atomic position and Zeeman sub-state, and can vary within a photocurrent segment because of complex atomic motion and imperfect optical polarization after many $\mu$s \cite{Dohe97,Kerc11a}, laser cooling and the cavity aperture ensures that atoms fall transversely through the mode in $\sim50\mu$s, stochastic heating should induce diffusion $<$100nm during the first 5$\mu$s of strong probing, and optical dipole forces should be minimal for near-maximally coupled atoms.  While technical noises like fluctuations in the atomic coupling during a transit, cavity instability and laser noise could also give rise to signals with super-shot noise variance, such sources may be ruled out by a combination of being too small to be of importance, of occurring only on significantly longer time scales, and/or the fact that they would induce super shot-noise distributions in both quadratures roughly equally due to our $\Theta\approx-\kappa$ probe detuning (such quadrature-independent fluctuations appear to be minimal in Fig.~\ref{fig:SS}, whereas the amplitude quadrature-only `quantum switching' is prominent).  Although the complexity of the measurements' dependence on $g$ precludes statistically rigorous parameter estimation, we find that a fixed effective value $g = 0.8\times g_0$ in our analysis provides a good visual fit to the distributions of these short-time datasets (corresponding to a near-optimal $\chi^2$ if one makes the gross simplification of fitting the experimental histograms to theoretical marginal distributions with fixed $g$, assuming that the uncertainties of the histogram data points are constant and uncorrelated) in all three parameter sets with $\sim 10\%$ uncertainty in $g$ and no other free parameters.  We believe that this approximation of fixed $g$ and the finite (20 MHz) bandwidth of our presented homodyne signals account for slight mismatches between theory and experiment in the amplitude quadrature splitting and phase-quadrature mean in the six data sets.  

\begin{figure}[tb]
\begin{center}
\includegraphics[width=0.6\textwidth]{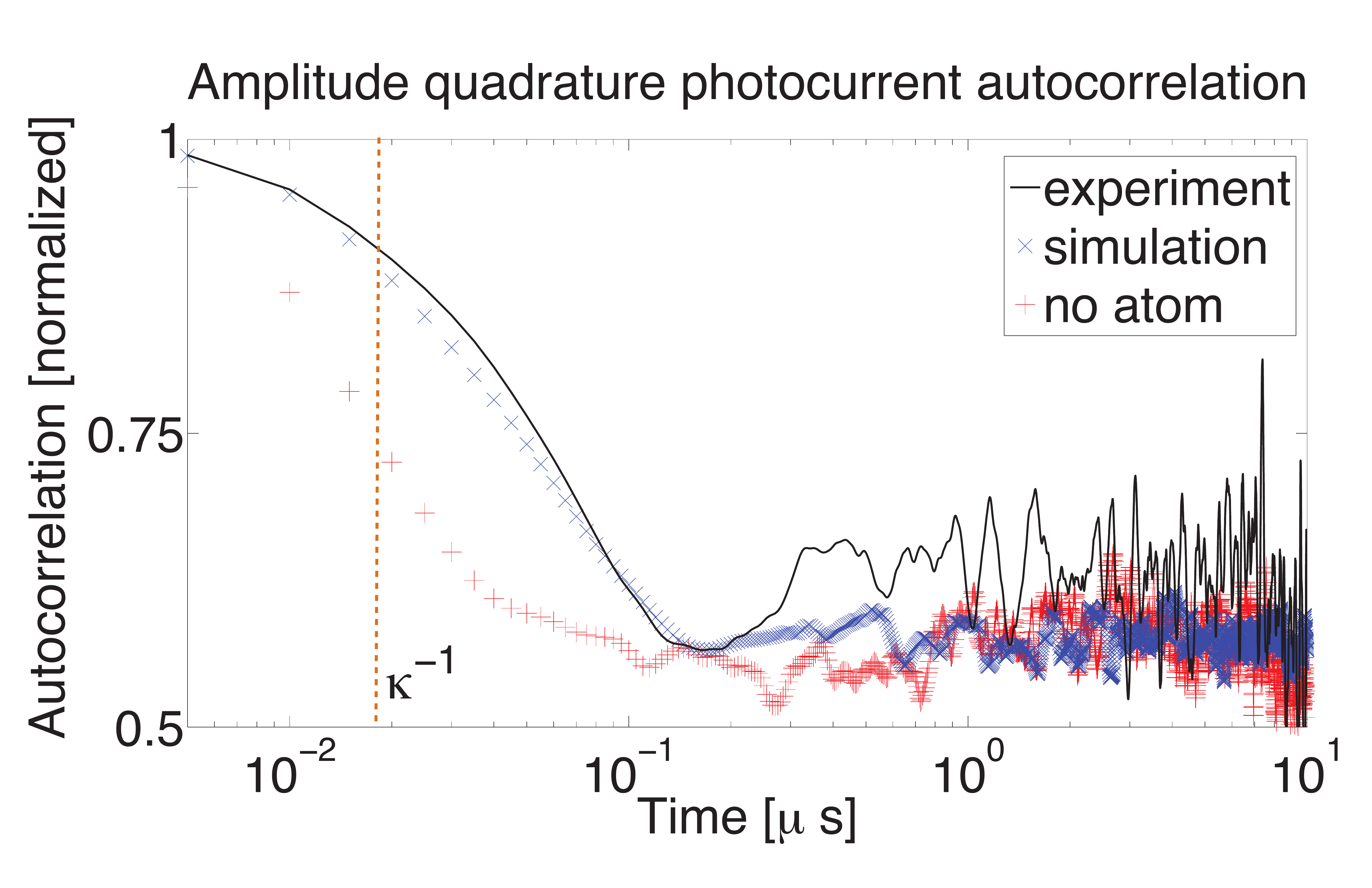}
\end{center}
\caption{\label{fig:corr}  The autocorrelation function for the same aggregated amplitude quadrature photocurrent data presented as a histogram in figure \ref{fig:SS}(d) is displayed in black.  Blue crosses represent the autocorrelation of photocurrents simulated by quantum trajectory methods \cite{Tan99} for identical parameters, as in Fig.~\ref{fig:Singles}(f). The stability of these quasi-bistable signals is enhanced relative to linearly scaled empty cavity transmission data taken after the atom is lost (red pluses, also presented with a 20MHz analog bandwidth and scaled to match the $0$ and $10\mu$s autocorrelation of the high-variance `experiment' signal), and the $\kappa^{-1}=17$ns cavity decay time (dashed orange line) characterizing the intracavity field relaxation of the empty resonator.  We attribute the elevated and noisy autocorrelation of the atom-cavity transmission data at $\gtrsim1\mu$s timescales to dynamic fluctuations in $g$, as in Fig \ref{fig:SS}.}
\vspace{-0.1in}
\end{figure}

\begin{figure}[tb]
\begin{center}
\includegraphics[width=0.77\textwidth]{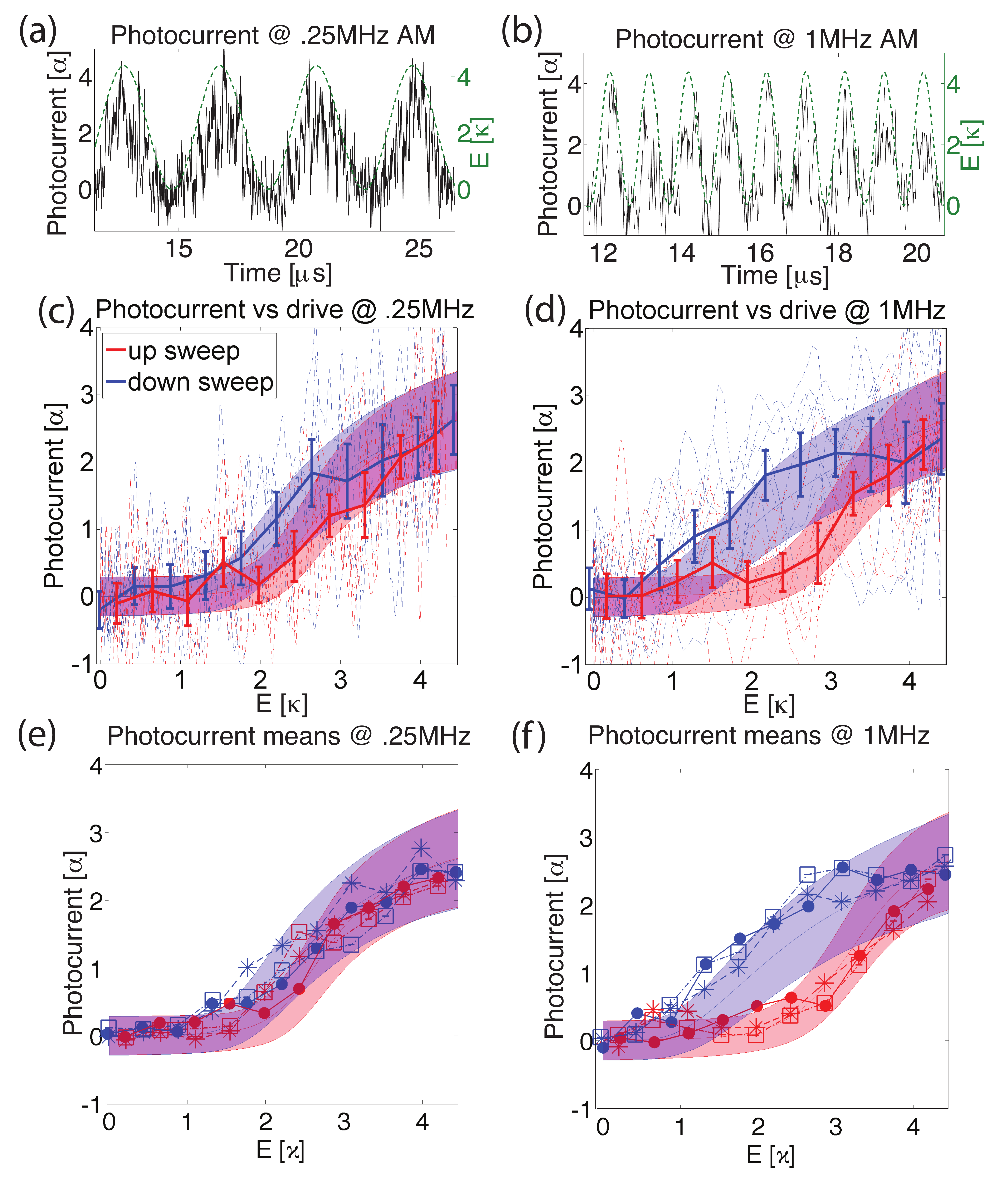}
\end{center}
\caption{\label{fig:AM}  (a-b) Single-shot, amplitude quadrature measurements as the drive amplitude is swept at .25MHz and 1MHz, respectively.  Black traces are 20MHz bandwidth photocurrents while the green, dashed traces represent the instantaneous drive amplitude.  (c-d) Dashed red (blue) traces portray the same photocurrent data in (a) and (b), respectively, as a function of the instantaneous, increasing (decreasing) drive amplitude.  Error bars represent sample mean and sample standard deviation of the same photocurrents within non-overlapping drive amplitude intervals.  Red and blue regions represent theoretically expected photocurrent mean and sample variance as a function of instantaneous drive (see Appendix \ref{sec:models}).  (e-f) Only the sample means of 3 additional single shot measurements of similar duration are plotted in each figure as a function of the instantaneous drive amplitude, as in (c) and (d), overlaying the same theoretically expected photocurrent statistics.}
\vspace{-0.1in}
\end{figure}

Hence, even in the single-atom, $\sim$10 photon regime, distinct high- and low-amplitude states of the output field are not washed out completely by quantum fluctuations. Remnant signatures of optical bistability are visible in the bimodal output photocurrents of Figs.~\ref{fig:Singles} and ~\ref{fig:SS}.  Similarly, as suggested in Fig.~\ref{fig:Singles}, we see in Fig.~\ref{fig:corr} that although the output field switches spontaneously when an atom is present in the cavity, it remains correlated over timescales much longer than that of light transmitted through an empty cavity. This atom-induced memory effect can be seen as an additional remnant of MBE-type optical bistability, where classically the high- and low-amplitude states are truly stable (due to negligible fluctuations in the system) and would therefore exhibit infinite correlation time. Consequently, it should also be possible to observe the hysteretic amplitude response characteristic of classical optical bistability by modulating the system drive slowly compared to the timescale for relaxation of the intracavity field (set in our case by the cavity decay time) but rapidly compared to the `metastable' memory timescale indicated in Fig.~\ref{fig:corr}. Accordingly, the data in Fig.~\ref{fig:AM} were obtained by recording amplitude-quadrature homodyne photocurrents while sweeping the drive strength sinusoidally at 0.25MHz or 1MHz.  Figs.~\ref{fig:AM}(a) and~\ref{fig:AM}(b) depict representative single-shot photocurrent segments with $\{\Theta,\Delta\} = \{-1.1,.7\}\kappa$ encompassing several cycles of sinusoidal drive amplitude modulation (AM) spanning the steady state bimodal region. Increases in both the mean and variance of the output photocurrent, largely in phase with the drive amplitude, can be discerned in both of these real-time plots. However, plotting the photocurrent as a function of the instantaneous drive amplitude (Figs.~\ref{fig:AM}(c) and~\ref{fig:AM}(d)) reveals a significant hysteresis in the system response at 1MHz AM that is barely noticeable at the more adiabatic .25MHz AM rate. Whereas the response of the empty cavity is linear and non-hysteretic with fixed (shot-noise) output photocurrent variance at these modulation frequencies, nonlinear increases in the signal mean and variance are evident in both traces at mid-sweep. At 1MHz AM, a hysteresis loop appears to open between the upward and downward drive amplitude sweeps, with the low (high) state persisting over a wider range of increasing (decreasing) drive amplitudes than at .25MHz AM.  Moreover, we emphasize that despite the large variance in the signals when viewed at such high bandwidth, the transition from a non-hysteretic to hysteretic mean response as the AM rate increases is significant, as highlighted in Figs.~\ref{fig:AM}(e-f).  These data are consistent with theoretical predictions (see Appendix~\ref{sec:models}), again assuming the same effective value of $g=.8\times g_0$ determined from the short-time segments analyzed in Fig.~\ref{fig:SS}. 

In conclusion, we have probed the dynamic optical response of a driven, strongly coupled, single-atom cQED system in the vicinity of atomic saturation. While our observations further confirm the predictions of a fully-quantum model, qualitative remnants of semiclassical absorptive bistability are clearly visible in the data. 
While the quantum fluctuation-limited lifetimes of the low- and high-amplitude states in attojoule optical `bistability' may be too short for direct use in photonic switching, our results clearly illustrate that existing theoretical models can be used to predict and analyze the dynamic response of real devices in the few photon regime. Such models can be used to make detailed predictions of the impact of quantum effects on ultra-low energy switch performance, which may be of interest to the nanophotonic engineering community. Beyond mere simulation and analysis, existing theoretical methods can be used to explore new approaches to the suppression of quantum fluctuations in the design of switches and related nanophotonic devices, for example by exploiting embedded coherent feedback control~\cite{Mabu11,Kerc10}. Even in a purely classical information processing paradigm, high spatial-density and ultra-low power nanophotonic circuit design presents intriguing new challenges for the nascent applied physics discipline of quantum engineering.

\section*{Appendix}
\appendix

\section{Experimental apparatus}\label{sec:exp}
\begin{figure}[tb!]
\begin{center}
\includegraphics[width=0.85\textwidth]{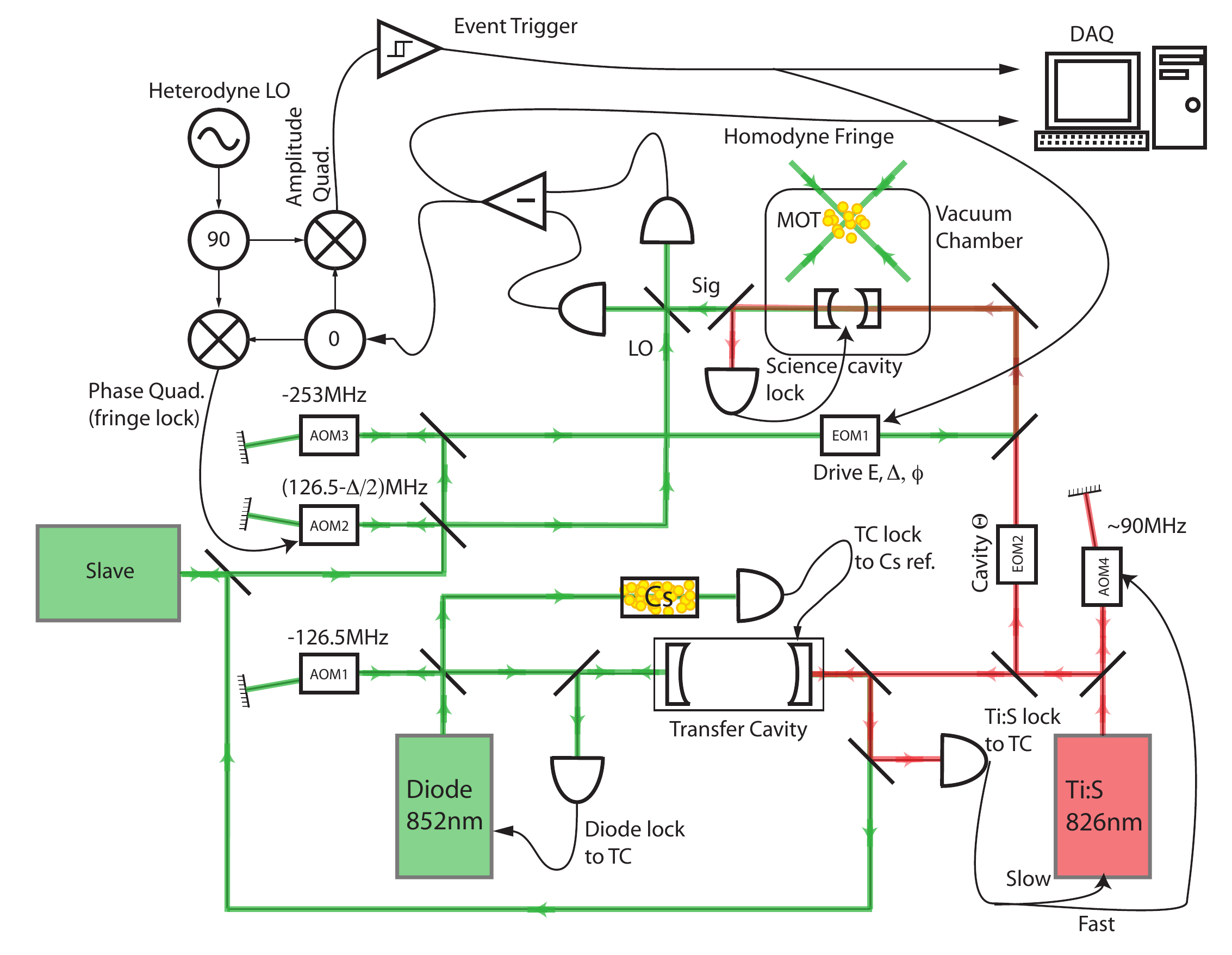}
\end{center}
\caption{\label{fig:exp} The essential optical and electro-optical components that define and stabilize the science cavity resonance frequency, drive the cQED system, detect individual atom transits, and perform homodyne detection of the cQED transmitted field, as explained in the text.}
\vspace{-0.1in}
\end{figure}

The experiment consists of a standard cQED setup involving laser cooled $^{133}$Cs atoms and a high finesse Fabry-Perot optical resonator \cite{Arme09,Kerc11a,Mabu96}.  Through frequency and polarization selectivity, we attempt to drive only the $\left(6S_{1/2}, F=4,m_F=+4\right)\rightarrow\left(6P_{3/2}, F=5, m_F=+5\right)$ atomic cycling transition at 852nm in order to approximate the atom as a two-level system.  Although the cavity used in the experiment was constructed to optimize signatures of spontaneous dressed-state polarization at high drive amplitudes \cite{Kerc11a}, it proved well-suited to study the amplitude `bistabiliy' dynamics presented in the main article.

Inside a UHV ($\approx10^{-9}$ Torr) chamber and placed on a multi-stage vibration-isolation stack, the Fabry-Perot optical resonator is formed by two high-reflectivity (8ppm transmission, 2ppm loss), 10cm radius of curvature dielectric mirrors with roughly 27$\mu$m of separation, yielding a 300,000-finesse optical resonator for the standing wave, TEM$_{00}$, 18$\mu$m-waist transverse spatial mode with a field decay rate of $\kappa = 2\pi\times9.3$MHz. We took particular care to mount the mirrors in a rotationally-symmetric manner to minimize stress-induced birefringence in the mirror coatings, allowing for full polarization-selectivity of the atomic transitions. The cavity length is tuned and actively stabilized by two shear-mode piezoelectric plates underlying the two mirror mounts. The precise cavity length and resonance frequency is continually stabilized by the Pound-Drever-Hall (PDH)~\cite{PDH} method using an additional laser probe detuned by the desired probe/cavity resonance frequency by two cavity free spectral ranges (at an optical wavelength of roughly 826nm, which interacts negligibly with Cs).

A Doppler-limited, magneto-optically trapped ensemble (MOT) of $\sim10^6$ atoms is formed roughly 1cm above the cavity mode in the UHV chamber. After cooling, the ensemble trap is switched off, allowing the cold atoms to fall under gravity towards the cavity mode and by the time they reach the cavity mode their free-fall velocity tends to dominate any residual thermal motion.  Due to the strong coupling between the targeted atomic transition and the cavity mode (with calculated maximum value $g_0=2\pi\times56.8$MHz at the cavity anti-nodes, using the dipole strength of the atomic transition and cavity mode volume), individual atom transits are detected by monitoring the ($g$-dependent) cavity transmission amplitude using a relatively weak and near-resonant probe \cite{Mabu96}, a free space balanced photodetector, and an actively phase-locked optical local oscillator (LO).  Although multiple atom transits per drop may be visible (apparent whenever $C\equiv g^2/2\kappa\gamma_\perp\gg1$), the atomic ensemble is sufficiently diffuse such that no more than one strongly-coupled atom is simultaneously present in the cavity mode and we acquire data from only one transit per ensemble drop.  It is in principle possible that a weakly coupled ``haze'' of background atoms are also coupled to the mode, but there is now a considerable literature (starting with \cite{Mabu96}) that demonstrates that dropped-atom cavity QED systems quantitatively reproduce the predictions of single-atom theory.  Once a strongly coupled atom has been detected, the probe power and frequency shift to the desired experimental levels and data acquisition is initiated.  

Fig. \ref{fig:exp} depicts many aspects of the resonance lock, optical drive, transmission measurement, and atom-triggering in the experiment (the laser cooling optical system is standard and remains tacit).  A diode laser is PDH frequency locked 253MHz to the blue of a large, mechanically stable `transfer' cavity mode, which is itself locked 253MHz to the red of the $\left(6S_{1/2},F=4\right)\rightarrow\left(6P_{3/2},F=5\right)$, 852nm hyperfine transition (so that the diode laser is on atomic resonance).  The purpose of this $\sim80,000$ finesse, 10kHz-linewidth transfer cavity is two-fold: provide a stable, Cs-locked frequency reference for all the lasers and act as a cleaning cavity for the diode laser, producing a more narrow linewidth 852nm laser source in transmission than is easily achievable in the laser lock.  The 852nm beam transmitted by the transfer cavity seeds a high power slave laser, which sources the light used in both the science cavity drive and optical local oscillator.  The `drive' arm of the slave output is frequency shifted to the red of the atomic transition again before entering an electro-optic modulator (EOM 1), which adds the optical carrier sideband near atomic resonance that serves as the science cavity's dynamic drive (the carrier and other sidebands are sufficiently detuned from the cQED system and have no measurable effect).  The science cavity is actively stabilized by a second, titanium-sapphire (Ti:S)  laser at roughly 826nm and frequency locked relative to the 852nm diode laser and the atomic transition via a frequency lock of the Ti:S laser to the transfer cavity: an optical sideband exactly two free spectral ranges to the red of the desired science cavity resonance near the atomic resonance is defined by a second EOM (EOM 2) and used to PDH resonance lock the science cavity in transmission.  The power of the locking laser was kept at $<$100nW so that the AC Stark shift it induced in the atoms could be neglected in our analysis.  As implied by the main article, the effects of the uncertainties in system parameters (e.g. laser noise, cavity resonance frequency instabilities, AC Stark shifts of the atom) are dominated by the effects of the variations in the atom-cavity coupling rate. 

For detecting and triggering off of strongly coupled atoms during their $\sim50\mu$s transit through the mode, the drive is initially tuned several MHz from the desired frequency for the experiment and its amplitude is set below the level of atomic saturation, while the several mW optical LO is tuned to the experimental drive frequency.  Both heterodyne quadratures of the transmitted atom-probe are detected via their interference with the optical LO.  The heterodyne phase quadrature is used to stabilize the relative phase of the free-space LO against slow drifts in the signal and LO path lengths via a 1kHz bandwidth optical phase lock loop.  The amplitude quadrature is monitored by a Schmitt trigger that fires when the measured field amplitude drops below a threshold indicating a strongly coupled atom in the cavity.  This trigger shifts the drive amplitude, frequency and LO-relative phase to the desired experimental configuration and initiates data acquisition of the homodyne photocurrent at 200MS/s.  Although collected at full bandwidth, the data presented in the main article has been additionally filtered for clarity using a 20MHz two-pole low pass, filtering out as much of the high frequency shotnoise as possible while still preserving the visibility of the fast, quantum fluctuations.

\section{cQED modeling and photocurrent predictions}\label{sec:models}

The driven Jaynes-Cummings Hamiltonian \cite{Berman,Carm93} is the standard quantum model of the internal dynamics of a two-level atom coupled to an optical resonator mode in a frame rotating with the external drive ($\hbar=1$)
\begin{equation}
H = \Delta\sigma^\dag\sigma+\Theta a^\dag a+ig(a^\dag\sigma-a\sigma^\dag)+ iE(a^\dag - a),
\end{equation}
where $a$ is the cavity mode annihilation operator, $\sigma$ is the atomic lowering operator and $^\dag$ denotes the Hermitian conjugate.  From left to right, the RHS terms correspond to the atom-drive detuning, the mode-drive detuning, the atom-mode coupling, and the external drive, respectively.  The complete quantum model of this system comes from extending these Hamiltonian dynamics to include processes associated with the dissipation of photons through the cavity mirrors at mean rate $2\kappa$ per intra-cavity photon and excited atomic state spontaneous emission at mean rate $2\gamma_\perp$, as modeled by $a$ and $\sigma$ operator-coupling to external quantum fields \cite{Berman, Carm93}.  Only one of these fields, the transmitted mode, is monitored by our homodyne detection setup.  For many applications, the entire model may be effectively represented by a \emph{master equation} that describes the unconditional evolution of any mode and/or atom operator $O$ (in the Heisenberg picture) \cite{QN,Carm93}
\begin{eqnarray}\label{eq:master}
\frac{d}{dt}O &=&  i[H,O] + 2\kappa\left(a^\dag Oa - \frac12a^\dag aO-\frac12Oa^\dag a\right) + 2\gamma_\perp\left(\sigma^\dag O\sigma - \frac12\sigma^\dag\sigma O-\frac12O\sigma^\dag\sigma\right)\nonumber\\
&\equiv&\mathcal{L}O.
\end{eqnarray}
For example, the (one dimensional) null space of the analogous, Schr\"{o}dinger picture `Liouvillian,' $\mathcal{L}_S$, corresponds to the steady state density matrix utilized in Figs.~\ref{fig:MF}(b) and \ref{fig:SS} in the main article.  This model may be trivially extended to describe the interaction of any number of two level atoms with the mode (or even with multiple modes).

Starting from this quantum model, the Maxwell-Bloch Equations (MBEs) may be derived by first assuming the approximation that atom-mode operator expectations factor \cite{Lugi84,Arme06}, {\it e.g.} $\langle a\sigma^\dag\rangle\approx\langle a\rangle\langle\sigma^\dag\rangle$ (see also \cite{Mabu08a}).  This approximation is equivalent to modeling the cavity mode as if it were a noiseless classical field coupling to a two-level atom or to an ensemble of atoms (as the case may be).  The MBEs can take the form of a set of five real, first-order, non-linear differential equations of motions for $\langle a\rangle$ and other expectations.  Steady state solutions for $\langle a\rangle$ (proportional to the mean field transmitted by a Fabry-Perot cavity) and their dynamical stability may be found.  Properly scaled, these mean-field solutions depend on the atom(s)-mode coupling only through a dimensionless parameter known as the `cooperativity,' $C = Ng^2/2\kappa\gamma_\perp$, where $N$ is the number of coupled atoms and $g$ the rate of coupling to a single atom \cite{Arme06}.  Thus, as long as the cooperativities are equal in both cases, this model predicts equivalent steady states for a coupled atom and for a coupled macroscopic atomic ensemble.  For highly non-linear, $C\gg1$ systems, however, the correlations between discrete excitations in the mode (photons) and the atom(s) that are ignored in the mean-field model may play a significant role in the overall dynamics when $N\sim1$, as demonstrated in the main article, with our experimental $C\lesssim$67, $N$=1 system.

Measurements of the external field transmitted by the cQED system are modeled using a quantum stochastic differential equation (QSDE) framework \cite{QN,Barc90}.  Loosely speaking, this model represents observables of the transmitted field measured by our homodyne detection setup as a linear combination of cavity mode observables and `quantum white noises.'  For example, expected photocurrent distributions and correlations may be calculated from the characteristic functional \cite{Barc90,QN}
\begin{equation}\label{eq:char}
\Phi_T[\beta_s] = \langle \mathcal{T}\exp\left\{\int_0^T\beta_sdB^\dag_{out,s}-\int_0^T\beta_s^\ast dB_{out,s}\right\}\rangle
\end{equation}
where $\beta_s$ is some complex-valued scalar function of time, $^\ast$ denotes complex conjugation, $B_{out,s}$ is the QSDE annihilation process of the measured, transmitted field, $\mathcal{T}$ is the time-ordering operator, and the expectation is taken over both the system and external field degrees of freedom.

Assuming a boxcar averaging photocurrent filter of width $\tau$ (so that $\beta_s=\beta$ for $0\leq s\leq\tau$ and  $\beta_s=0$ otherwise), the Fourier transform of Eq.~\eqref{eq:char} produces a  representation of the field analogous to a Wigner quasi-probability distribution
\begin{equation}\label{eq:Wfn}
W_t(\alpha) = \frac{1}{\pi^2}\int d^2\beta e^{\alpha\beta^\ast-\alpha^\ast\beta}\text{Tr}\left[ e^{\tau(\mathcal{L}-\frac12\vert\beta\vert^2+\mathcal{M})}\rho_t\right]
\end{equation}
where $\rho_t$ is the reduced density matrix for the cQED system alone at time $t$ (may be the steady state density matrix, for example), $\mathcal{M}\rho_t = \beta\sqrt{2\kappa\eta}\rho_ta^\dag- \beta^\ast\sqrt{2\kappa\eta}a\rho_t$ is a `measurement superoperator,' and $\eta$ is the efficiency with which photons decaying from the cavity are measured by the homodyne detector.  As with a Wigner function, $P_t(\alpha_\theta)$, the probability that a homodyne measurement of quadrature $\theta$ at time $t$ will take the (real) value $\alpha_\theta$, is obtained by integrating $W_t(\alpha)$ over the complementary quadrature, 
\begin{equation}\label{eq:Marginal}
P_t(\alpha_\theta) = \int d\alpha_{\theta+\pi/2}W_t(\alpha).
\end{equation} 
In practice, we approximate $W_t(\alpha)$ by assuming short photocurrent integration times so that we may `freeze' the relevant internal dynamics over the sample interval, taking $\tau\mathcal{L}\rightarrow 0$.  This approximation also implies the assumption that the Gaussian fluctuations in the input vacuum field are not correlated with system state over the effective measurement time interval.  As a result, in this approximation $W_t(\alpha)$ is equivalent to a convolution of the Wigner function for the system state and a symmetric, mean-zero Gaussian distribution with a variance set by the effective integration time (which represents the contribution of vacuum field fluctuations to the measurement distribution).  These methods and approximations were employed in Fig.~\ref{fig:SS} of the main article. 

Similarly, integrating Eq.~\eqref{eq:char} over some $\beta_{\theta+\pi/2}$-quadrature produces the characteristic functional
\begin{equation}
\Phi_{\theta,T}[k] = \langle \mathcal{T}\exp\left\{i\int_0^Tk_{s}dY^\dag_{\theta,s}\right\}\rangle,
\end{equation}
where $k_{s}$ is an aribtrary real-valued scalar function of time and $dY_{\theta,s}/dt\equiv I_{s}$ is the homodyne photocurrent operator of quadrature $\theta$.  This functional may be used to calculate moments of instantaneous photocurrent measurements \cite{Barc90,QN}
\begin{equation}\label{eq:moments}
\langle I_{t_1}...I_{t_n}\rangle = (-i)^n\frac{\partial^n}{\partial k_{t_1}...\partial k_{t_n}}\Phi_{\theta,T}[k]\vert_{k=0}.
\end{equation}
For instance, the mean amplitude-quadrature photocurrent at time $t$ is calculated to be
\begin{equation}\label{eq:1time}
\langle I_t\rangle = {\text Tr}\left[-i\mathcal{M}_0\rho_t\right]
\end{equation}
where $-i\mathcal{M}_0\rho_t = \sqrt{2\kappa\eta}\left(\rho_ta^\dag + a\rho_t\right)$, while the two-time photocurrent correlation is ($t<t^\prime<T$)
\begin{equation}\label{eq:2time}
\langle I_{t^\prime}I_t\rangle = \delta(t-t^\prime)-{\text Tr}\left[\mathcal{M}_0\exp\{\mathcal{L}(t^\prime-t)\}\mathcal{M}_0\rho_t\right].
\end{equation}
Note that the first term on the RHS of Eq.~\eqref{eq:2time} may be identified as the shotnoise contribution to the photocurrent correlation function, while the second is the contribution from the system.  As the detector-filtered photocurrent operator is $\mathcal{I}_t=\int_0^Tf_t(s)I_sds$ where $f_t(s)$ is some filter function imposed by the detection at time $t$ (e.g. a low-pass filter initiated at time $t$), the above two equations may in principle be used to calculate the mean and variance of filtered photocurrent measurements.  In practice, though, we again invoke a small integration time approximation, justified by our high bandwidth detection.  If we approximate the state $\rho_t$ as static over the effective integration time (as far as detection is concerned), we may take $\rho_t$ in the RHS of Eqs.~\eqref{eq:1time} and~\eqref{eq:2time} as independent of time over the effective integration interval and approximate $\mathcal{L}(t^\prime-t)\approx0$, greatly simplifying these calculations.  These approximations again allow us to model the detector filter function by a simple time-averaging filter of width $\tau=2/(\pi f_c)$, where $f_c=20$MHz is the cut-off frequency of the two-pole low pass filter that sets the bandwidth of the data presented in the main article.  Thus the mean and variance of the photocurrent measurements expected from our detector when the system is in the state $\rho_t$ may be calculated from 
\begin{eqnarray}\label{eq:Is}
\langle\mathcal{I}_t\rangle &=& \tau{\text Tr}\left[-i\mathcal{M}_0\rho_t\right]\nonumber\\
\langle\mathcal{I}_t^2\rangle &=& \tau -\tau^2{\text Tr}\left[ \mathcal{M}_0^2\rho_t\right]
\end{eqnarray}
These methods and approximations were used to calculate the theory curves in Fig.~\ref{fig:AM} of the main article.  Integrating the Schrodinger-picture master equation for AM drive (initiating system and drive in the ground state) returns an expected ensemble system state $\rho_t$ at each point in a drive cycle.  These states were then used to calculate the expected ensemble mean and standard deviation of the photocurrent measurements using Eqs.~\eqref{eq:Is}, given the bandwidth of the data.  Relating the master equation's quantum ensemble, single cycle predictions to the several cycle, single shot data in Fig.~\ref{fig:AM} is justified by an assumption of the system's ergodicity (see below). 

Finally, we relied on quantum trajectory methods \cite{Carm93}, which compliment the above master equation-based approaches to simulate typical measurements expected from our apparatus in Figs.~\ref{fig:Singles} and~\ref{fig:corr}.  While the master equation may be used to model ensembles of experimental realizations, a simulated quantum trajectory may be used to construct potential experimental homodyne measurement sequences, correctly sampled from the space of all possible sequences.  For example, simulation of an amplitude quadrature photocurrent given a set of experimental parameters first involves the calculation of a possible trajectory for the internal quantum state vector $\vert\psi_c(t)\rangle$ by numerically integrating the stochastic Schrodinger equation~\cite{Tan99,Carm93}
\begin{eqnarray}
d\vert\psi_c(t)\rangle &=& -(iH+\kappa a^\dag a+\gamma_\perp\sigma^\dag\sigma)\vert\psi_c(t)\rangle dt+\left(\sqrt{2\kappa}\langle a+a^\dag\rangle_c dt+dW_t^{(1)}\right)\sqrt{2\kappa}a\vert\psi_c(t)\rangle+\nonumber\\
&&\left(\sqrt{2\gamma_\perp}\langle\sigma+\sigma^\dag\rangle_c dt+dW_t^{(2)}\right)\sqrt{2\gamma_\perp}\sigma\vert\psi_c(t)\rangle
\end{eqnarray}
where $\{dW_t^{(1)},dW_t^{(2)}\}$ are randomly generated, independent Wiener increments, $\langle\cdot\rangle_c$ denotes expectation with respect to $\vert\psi_c(t)\rangle$ and the state vector is forcibly re-normalized after each recursive update.  The simulated photo-increment $dY_t$ may then be obtained using this state trajectory and calibrated detection efficiency $\eta$
by
\begin{equation}
dY_t = \sqrt{\eta}\left(\sqrt{2\kappa}\langle a+a^\dag\rangle_c dt+dW_t^{(1)}\right)+\sqrt{1-\eta}dW_t^{(3)}
\end{equation}
where $dW_t^{(3)}$ is a third, independent Wiener increment.

\begin{figure}[tb!]
\begin{center}
\includegraphics[width=0.85\textwidth]{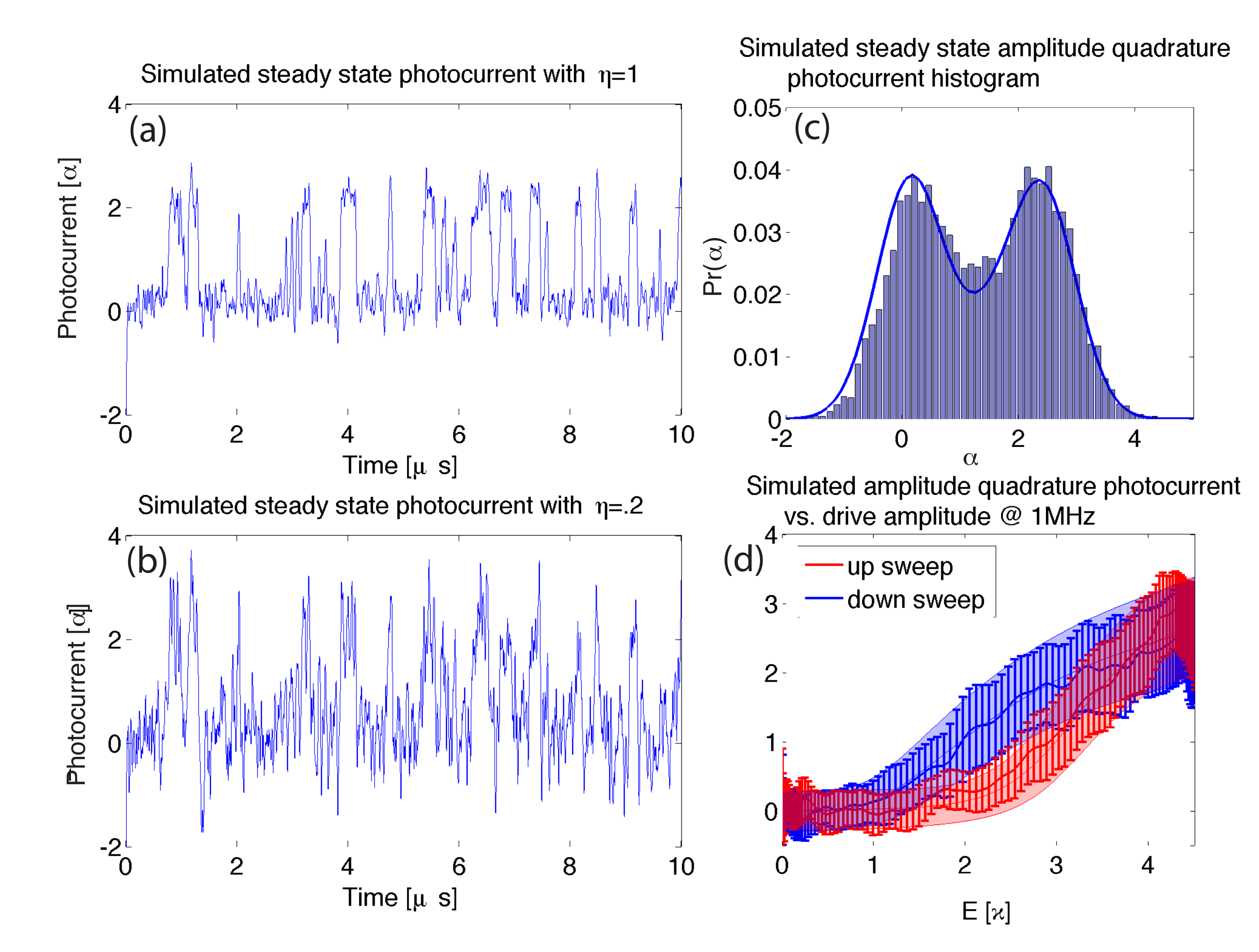}
\end{center}
\caption{\label{fig:sims} Reproduction of several figures in the main article replacing experimental data with simulated photocurrent data produced by quantum trajectory methods.  (a), Simulated amplitude quadrature photocurrent data with $\{\Theta,\Delta,E\} = \{-1.1,.7,2.6\}\kappa$, 20MHz analog bandwidth and perfect detection efficiency.  (b), Same quantum trajectory realization as (a), but with the 20MHz bandwidth photocurrent simulated with $\eta=.2$ efficiency.  (c), Histogram of simulated inefficient photocurrents in comparison to the expected distribution derived from master equation based techniques.  (d), Sample mean and sample standard deviation of simulated photocurrents from fifty 1MHz AM cycles, overlaying master equation-based predictions.}
\vspace{-0.1in}
\end{figure}

Again, quantum trajectory simulations more directly reproduce time-series, experimental measurements, while master equation-based simulations represent ensembles of trajectories or measurements.  In Fig. \ref{fig:sims} we present many of the same figures in the main article using simulated photocurrent data based on quantum trajectories in comparison with the master equation predictions.  Figs. \ref{fig:sims}(a-b) depict simulated amplitude quadrature photocurrents with the same experimental parameters as in Figs.~\ref{fig:MF}(c),~\ref{fig:Singles} and~\ref{fig:SS}(d) in the main article.  In Fig. \ref{fig:sims}(a), the measurement of the the mode leaking from the cavity is simulated as detected with perfect efficiency, while Fig. \ref{fig:sims}(b) simulates the same realization, but with our experimentally typical 20\% efficiency, as in Fig.~\ref{fig:Singles}(e-f).  Fig. \ref{fig:sims}(c) histograms a simulated inefficient photocurrent trajectory in comparison with the expected amplitude quadrature distribution using Eq.~\eqref{eq:Marginal} ({\it i.e.} reproducing the amplitude quadrature component of Fig.~\ref{fig:SS}(d) in the main article using simulated data).  Finally, Fig. \ref{fig:sims}(d) compares the sample mean and sample standard deviation of simulated amplitude quadrature measurements during 50 cycles of a 1MHz AM external drive, for comparison to Fig.~\ref{fig:AM}(d) in the main article.  Both Figs. \ref{fig:sims}(c) and (d) utilize several times more aggregate data than the experimental data presented in the main article in order to articulate the convergence of predictions from quantum trajectories to those of master equation simulations.  As we expect both types of systems to be stationary ergodic processes, we have empirically confirmed that the main discrepancies between quantum trajectory and steady state master equation based simulations observed in Fig. \ref{fig:sims} arise from the marginal appropriateness of the $\tau\mathcal{L}\rightarrow0$ approximation when modeling 20MHz bandwidth signals derived from our system.

\section*{Acknowledgments}
This work is supported by DARPA-MTO under Award No.\ FA8650-10-1-7007.


\begin{thebibliography}{99}

\bibitem{Obri09} J. L. O'Brien, A. Furusawa, and J. Vu\v{c}kovi\'{c}, ``Photonic quantum technologies,'' Nature Photonics {\bf 3}, 687-695 (2009).

\bibitem{Aoki09} T. Aoki, A. S. Parkins, D. J. Alton, C. A. Regal, B. Dayan, E. Ostby, K. J. Vahala, and H. J. Kimble, ``Efficient Routing of Single Photons by One Atom and a Microtoroidal Cavity,'' Phys. Rev. Lett. {\bf 102}, 083601 (2009).

\bibitem{DiCa10} L. DiCarlo, M. D. Reed, L. Sun, B. R. Johnson, J. M. Chow, J. M. Gambetta, L. Frunzio, S. M. Girvin, M. H. Devoret, and R. J. Schoelkopf, ``Preparation and measurement of three-qubit entanglement in a superconducting circuit,'' Nature {\bf 467}, 574-578 (2010).

\bibitem{Sava88} C.~Savage and H.~.J.~Carmichael, ``Single-atom optical bistability,'' IEEE J.\ Quantum Electron.\ {\bf 24}, 1495-1498 (1988).

\bibitem{Remp91} G. Rempe, R. J. Thompson, R. J. Brecha, W. D. Lee, and H. J. Kimble, ``Optical bistability and photon statistics in cavity quantum electrodynamics,'' Phys. Rev. Lett. {\bf 67}, 1727 (1991).

\bibitem{Arme06} M.~A.~Armen and H.~Mabuchi, ``Low-lying bifurcations in cavity quantum electrodynamics,'' Phys.\ Rev.\ A {\bf 73}, 063801 (2006).

\bibitem{Kerc11a} J. Kerckhoff, M. A. Armen, D. S. Pavlichin, and H. Mabuchi, ``The dressed atom as binary phase modulator: towards attojoule/edge optical phase-shift keying,'' Opt. Express {\bf 19}, 6478-6486 (2011).

\bibitem{Mill10} D. A. B. Miller, ``Are optical transistors the logical next step?,'' Nature Photonics {\bf 4} 3-5 (2010).

\bibitem{Stro01} S. H. Strogatz, {\it Nonlinear Dynamics and Chaos: With Applications to Physics, Biology, Chemistry and Engineering} (Perseus, Cambridge, MA, 1994).

\bibitem{Mabu11} H.~Mabuchi, ``Coherent-feedback control strategy to suppress spontaneous switching in ultralow power optical bistability,'' Appl. Phys. Lett. {\bf 98}, 193109 (2011).

\bibitem{Srin07} K.~Srinivasan and O.~Painter, ``Linear and nonlinear optical spectroscopy of a strongly coupled microdisk-quantum dot system,'' Nature {\bf 450}, 862-865 (2007).

\bibitem{Fara08} A.~Faraon, I.~Fushman, D.~Englund, N.~Stoltz, P.~Petroff, and J. Vu\v{c}kovi\'{c}, ``Coherent generation of non-classical light on a chip via photon-induced tunnelling and blockade,'' Nature Physics {\bf 4}, 859-863 (2008).

\bibitem{Lugi84} L.~A.~Lugiato, in {\it Progress in Optics}, edited by E.~Wolf (North-Holland, Amsterdam, 1984), Vol.\ XXI.

\bibitem{Szok69} A. Sz\"{o}ke, V. Daneu, J. Goldhar, and N. A. Kurnit, ``Bistable optical element and its applications,'' Appl. Phys. Lett. {\bf 15}, 376 (1969).

\bibitem{Smit86} S.~D.~Smith, ``Optical bistability, photonic logic, and optical computation,'' Appl.\ Opt.\ {\bf 25}, 1550-1564 (1986).

\bibitem{Kili91} S. Ya. Kilin and T. B. Krinitskaya, ``Single-atom phase bistability in a fundamental model of quantum optics,'' J. Opt. Soc. Am. B {\bf 8}, 2289 (1991).

\bibitem{Hood98} C. J. Hood, M. S. Chapman, T. W. Lynn, and H. J. Kimble, ``Real-time cavity QED with single atoms,'' Phys.\ Rev.\ Lett.\ {\bf 80}, 4157-4160 (1998).

\bibitem{Yang07} X.~Yang, C.~Husko, C.~W.~Wong, M.~Yu, and D.~L.~Kwong, ``Observation of femtojoule optical bistability involving Fano resonances in high-Q/Vm silicon photonic crystal nanocavities,'' Appl.\ Phys.\ Lett.\  {\bf 91}, 051113 (2007).

\bibitem{Noza10} K. Nozaki, T. Tanabe, A. Shinya, S. Matsuo, T. Sato, H. Taniyama, and M. Notomi, ``Sub-femtojoule all-optical switching using a photonic-crystal nanocavity,'' Nature Photonics {\bf 4}, 477-483 (2010).

\bibitem{Carm86} H.~J.~Carmichael in {\it Frontiers in Quantum Optics}, edited by E.~R.~Pike and S.~Sarkar (Adam Hilger, Bristol, 1986).

\bibitem{Gang90a} H.~Gang, C.~Z.~Ning and H.~Haken, ``Codimension-two bifurcations in single-mode optical bistable systems,'' Phys.\ Rev.\ A {\bf 41}, 2702 (1990). 

\bibitem{Gang90b} H.~Gang, C.~Z.~Ning and H.~Haken, ``Distribution of subcritical Hopf bifurcations and regular and chaotic attractors in optical bistable systems,'' Phys.\ Rev.\ A {\bf 41}, 3975 (1990).

\bibitem{Mabu08a} H.~Mabuchi, ``Derivation of Maxwell-Bloch-type equations by projection of quantum models,'' Phys.\ Rev.\ A {\bf 78}, 015801 (2008).

\bibitem{Arme09} M. A. Armen, A. E. Miller, and H. Mabuchi, ``Spontaneous dressed-state polarization in the strong driving regime of cavity QED,'' Phys. Rev. Lett. {\bf 103} 173601 (2009).

\bibitem{Dohe97} A. C. Doherty, A. S. Parkins, S. M. Tan, and D. F. Walls, ``Motion of a two-level atom in an optical cavity,'' Phys. Rev. A {\bf 56}, 833 (1997).

\bibitem{Tan99} S.~M.~Tan, ``A computational toolbox for quantum and atomic optics,'' J.\ Opt.\ B: Quantum Semiclass.\ Opt.\ {\bf 1}, 424--432 (1999).

\bibitem{Kerc10} J. Kerckhoff, H. I. Nurdin, D. S. Pavlichin, and H. Mabuchi, ``Designing quantum memories with embedded control: Photonic circuits for autonomous quantum error correction,'' Phys. Rev. Lett. {\bf 105} 040502 (2010).

\bibitem{Mabu96} H.~Mabuchi, Q.~A.~Turchette, M.~S.~Chapman, and H.~.J.~Kimble, ``Real-time detection of individual atoms falling through a high-finesse optical cavity,'' Opt.\ Lett. {\bf 21}, 1393--1395 (1996).

\bibitem{PDH} R.~W.~P.~Drever, J.~L.~Hall, F.~V.~Kowalski, J.~Hough, G.~M.~Ford, A.~J.~Munley, H.~Ward, ``Laser phase and frequency stabilization using an optical resonator,'' Appl. Phys. B {\bf 31} 97 (1983).

\bibitem{Berman} P.~Berman., Ed.,  {\it Cavity Quantum Electrodynamics} (San Diego: Academic Press, 1994).

\bibitem{Carm93} H.~J.~Carmichael, {\it An Open Systems Approach to Quantum Optics} (Springer, Berlin, 1993).

\bibitem{QN} C.~W.~Gardiner and P.~ Zoller, {\it Quantum Noise} (Springer-Verlag, Berlin, 2004).

\bibitem{Barc90} A.~Barchielli, ``Direct and heterodyne detection and other applications of quantum stochastic calculus to quantum optics,'' Quantum Opt. {\bf 2} 423-441 (1990).

\end{thebibliography}
\end{document}